# High-Speed Train Cell-less Network Enabled by XGS-PON and Impacts on vRAN Split Interface Transmission

A. El Ankouri[1,2], L. Anet Neto[1], G. Simon[1], H. Le Bras[1], A. Sanhaji[1], P. Chanclou[1]
*(1) Orange Labs, 2 Avenue Pierre Marzin - 22300 Lannion, France, anas.elankouri@orange.com*
*(2) IMT Atlantique, 655 Avenue du Technopôle, 29200 Plouzané*

**Abstract:** We successfully demonstrate a transmission of a high layer split mobile interface for cell-less, high-speed train network applications using a commercially available XGS-PON. Operation is also demonstrated for a GbE interface.

## 1. Introduction

Providing higher bit-rates all the time and anywhere is a key feature of the coming 5G Radio Access Network (RAN). This includes solutions targeting vehicular transportation, which, of course, should allow for a seamless connectivity for the final users. As far as rail transportation is concerned, GSM-R [1] relies entirely on circuit switching and can only be used for voice communication and signaling operations for trains and railways. For the coming mobile generation, the requirements for a successor railway communication technology beyond 3GPP's Future Railway Mobile Communication System (FRMCS) are given in TS 22.289 [2].

In this context, efforts focusing on extending the mobile coverage in rural areas are currently being intensified to improve mobile services in high-speed trains (HST). Highly densely populated urban areas also represent a challenging use-case, where often poorly covered deep-indoor infrastructures such as undergrounds should also be addressed. Whereas signal penetration remains the major issue for deep-indoor scenarios, guaranteeing high data rates can be challenging in HST due to the speeds at which users are moving, generally beyond 300 km/h. Indeed, with fast mobility comes great complexity. The management of the mobile handover as well as the propagation limitations imposed by the Doppler frequency shift [3] need thus to be taken into consideration. A way to overcome the challenges of high-speed handover has been recently proposed where a mobile "cell-less" configuration is obtained thanks to an optical transmission system. A successful transmission was demonstrated in [4] using Analog Radio-over-Fiber (A-RoF) and per-antenna wavelength allocation. Wavelength Division Multiplexing (WDM) handover-free operation was enabled by a fast tunable laser with service interruption times in the order of some µs.

Here, we propose an equivalent cell-less network which is enabled, differently, by a commercially available 10 Gb/s symmetrical Passive Optical Network (XGS-PON) system. In our work, each Trackside Remote Unit (TRU) is connected to an Optical Network Unit (ONU), the latter being connected to the Optical Line Termination (OLT) through a classical point-to-multipoint (PtMP) topology as shown in Fig. 1. The PON will actually behave most the time as a point-to-point (PtP) system, being only temporarily switched to a 1:2 PtMP transmission during the transitions between an ONU and its neighbor as the HST moves along the rail. This solution would notably allow us to keep simple and robust to optical noise on-off keying (OOK) operation with low-cost commercially available off-the-shelf (COTS) small factor plug-and-play optical transceivers (SFP+) with a standardized and soon to be deployed optical access solution. First, we experimentally demonstrate the principle of the proposed solution

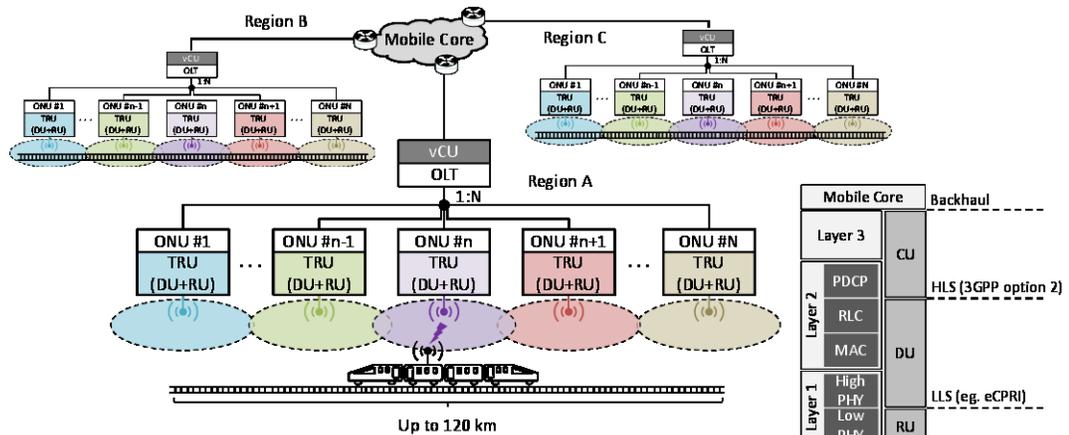

Fig. 1: HST XGS-PON "cell-less" network.

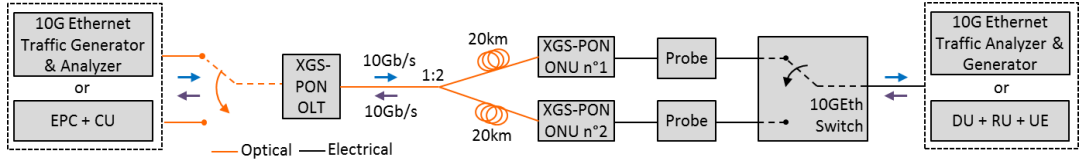
Fig. 2: Experimental Setup

and the switching between two ONUs with an Ethernet interface emulating a RAN functional split interface [5]. Then, we evaluate the impacts of the transmission of a real-time high-layer split (HLS) virtual RAN (vRAN) interface [6]. Finally, we address the XGS-PON parameters that would allow transmission at different RF frequencies.

## 2. Experimental setup

The experimental setup of our XGS-PON capable of enabling bi-directional fast switching between ONUs is depicted in Fig. 2. The top left-most block is an Ethernet traffic generator that emulates any mobile functional split option. We evaluate a GbE interface. Apart from this generic Ethernet interface, we also assess a real-time packet data convergence protocol/radio link control (PDCP/RLC) functional split corresponding to 3GPPs option 2 (c.f. inset Fig 1). This solution is based on virtualized central units (vCU) running on an Openstack cloud platform (for more details, please refer to [6]). This traffic is fed to an XGS-PON OLT connected to two ONUs through an optical coupler. The ONUs are connected to a probe device that allow us to visualize the bit-rate evolution and switching in time for our PON system.

The switching between TRUs is emulated by a 10GbE switch. The switching method relies on the aging time of its media access control (MAC) address table which is configured to 20s. Connected simultaneously to both of our ONUs, after 20s the switch transmits the flow from one ONU to the other, similar to what would happen on an actual transition of a train from one TRU to the next one. Finally, an Ethernet packet analyzer is used to evaluate the transmission of the generic GbE interfaces. A distributed unit (DU) and an emulated radio unit (RU) and user equipment (UE) are used to complete the downlink transmission for the PDCP/RLC split scenario. Our setup also supports upstream traffic with standard time-division multiple access (TDMA) operation. It should be noted that the main advantage of a virtualized CU in our approach is that it could facilitate the implementation of longer reach railways. Indeed, the handover of the mobile backhaul interface between zones covered by different PON trees could be easily implemented by simple transfer of vCU virtual machines between regional nodes (see Fig. 1).

In order to demonstrate the feasibility of the cell-less network with an XGS-PON system, the degradations during the transitions from on ONU TRU to another should be minimized. To do so, we explicitly disable anti-spoofing rules of our PON during a short period of time in which the two ONUs will be transmitting at the same time in the uplink with the same source MAC address. In a real implementation, some sort of white list police would have to be implemented for groups of three consecutive ONUs (previous/current/next) to allow for localized MAC flapping.

## 3. Results

The first set of results corresponds to the PDCP-RLC real-time mobile transmission. We transmitted a stream of 1200 bytes user datagram protocol (UDP) packets at a 100 Mbit/s. We assigned to our mobile traffic a dynamic bandwidth allocation (DBA) profile using a type 2 T-CONT with 150 Mbit/s fixed bandwidth to assure that our

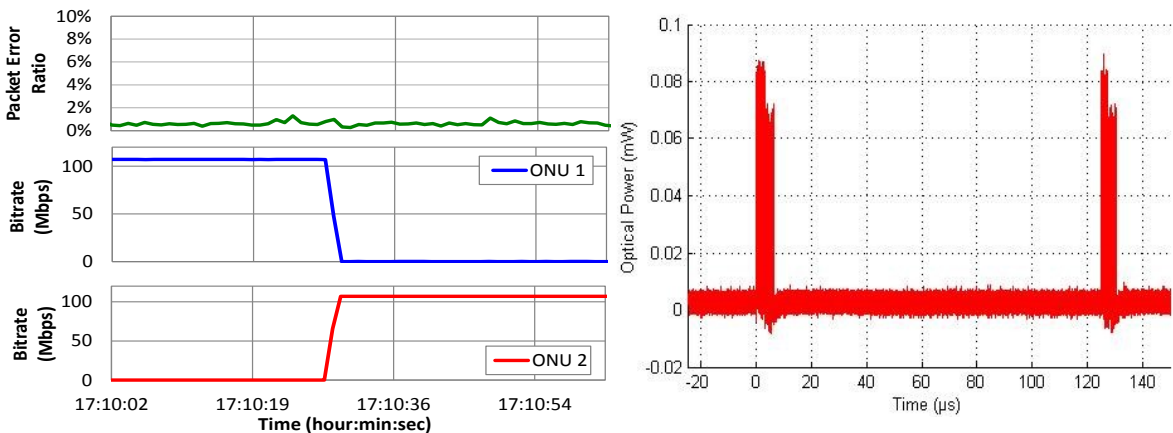
Fig. 3: ONUs bitrate and PER evaluations (left) and OLT bursts corresponding to the two ONUs in the uplink (right).

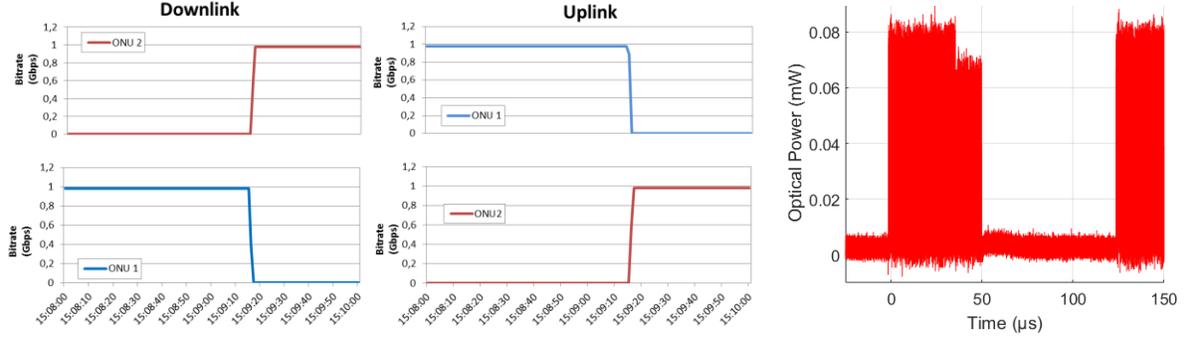

Fig. 4: ONUs bit-rate variation for the down and uplinks (left) and uplink bursts (right).

system could handle eventual bandwidth bursts higher than 100 Mbit/s without having to discard packets. In Fig. 3 (left), at the 17:10:28 time mark, the switch allows for a transition from ONU 1 to ONU 2. The transition period had a value of three seconds in which the communication wasn't interrupted and our system didn't suffer any additional Packet Errors at the reception of our mobile date. We should notice that this time limitation is exclusively due to our railway emulation. On an actual implementation, switching times could possibly be as small as 125μs, which corresponds to the minimum duty cycle and frame duration on a PON system. Fig.3 (right) we can two cycles of the upstream transmission with a higher power burst corresponding to ONU 1.

Fig. 4 shows the results for the generic GbE interface. In Fig. 4 (left), we can see that at 15:09:16, for both uplink and downlink transmissions, the transition from ONU 1 to ONU 2 is done under less than 1 second. During this time interval, both ONUs are transmitting at the same time. The DBA profile in this case was configured to transmit a maximum bandwidth up to 9 Gbit/s with an assured bandwidth of 1 Gbit/s. We didn't stress our system to the limit due to a bandwidth limitation of our probe device, incapable of providing throughput visualization beyond 1 Gbit/s. Fig. 4 (right) shows the two optical bursts of the ONUs.

## 4. Conclusions and possible deployment scenarios

In this work, we demonstrated a real-time bi-directional transmission of a generic functional split GbE interface as well as a PDCP-RLC split at 150 Mbit/s in an emulated HST scenario. Our cell-less network is demonstrated using a commercial XGS-PON system allowing operation with low-cost COTS SFP+ transceivers and standard OOK operation. In order to open up the possibilities of such solution, Tab. 1 shows three possible deployment use-cases. For instance, for a 700 MHz carrier, an 1:4 PON compatible with optical budget class N2 could be considered. By setting the distance between adjacent ONUs in the railway to 40 km, we would allow compatibility with the maximum differential distance between ONUs and between ONU and OLT [7] while guaranteeing a cell radius below 6.5 km (from Friis equation [8]). This would allow for an uninterrupted communication time of approximately 26 minutes per PON tree considering a speed of 360 km/h.

Tab. 1: Three use-cases for cell-less XGS-PON.

| Carrier frequency | Max. cell radius (Friis)[1] | #ONUs | $\Delta L_{ONU}$ \| cell radius[2] | $\Delta L_{DIFF., MAX}$[3] | $L_{MAX}$[4] | OB[5] | T[6] |
|---|---|---|---|---|---|---|---|
| 700 MHz | 34 km | 4 | 40 km \| 20 km | 40 km | 60 km | 31 dB (N2) | 26 min |
| 3.5 GHz | 6.5 km | 16 | 5 km \| 2.5 km | 35 km | 37.5 km | 29 dB (N1) | 13 min |
| 25 GHz | 1 km | 64 | 1 km \| 0.5 km | 31 km | 31.5 km | 36 dB (E1) | 10 min |

[1] For an RF link budget of 120dB  
[2] $\Delta L_{ONU}$: distance between adjacent ONUs on a rail track  
[3] $\Delta L_{DIFF., MAX}$: maximum differential distance between ONUs  
[4] $L_{MAX}$: Maximum distance between ONU and OLT  
[5] Minimum needed optical budget/class  
[6] Uninterrupted communication time per PON (360 km/h)